\documentclass[aps,twocolumn,superscriptaddress,groupedaddress,longbibliography,nofootinbib]{revtex4-2}
\usepackage{balance}
\usepackage{graphicx}  
\usepackage{dcolumn}   
\usepackage{bm}        
\usepackage{amssymb}   
\usepackage{amsmath}
\usepackage{amsfonts,amsthm}
\usepackage[english]{babel}
\usepackage[utf8]{inputenc}
\usepackage{indentfirst}
\usepackage{slashed}
\usepackage{mathrsfs}
\usepackage{amssymb}
\usepackage{color}
\usepackage{mathrsfs}
\usepackage{array}
\usepackage{subcaption}
\usepackage{float}

\usepackage{verbatim}

\mathchardef\mhyphen="2D

\begin{document}

\title{AdS$_9$ solutions in type II supergravities}
\author{Giuseppe Dibitetto$^{a, b}$ and Nicolò Petri$^c$\vspace{2mm}\\
\emph{$\mbox{}^a$Dipartimento di Fisica, Università di Roma “Tor Vergata", Via della Ricerca Scientifica 1, 00133,
Roma, Italy.\vspace{0.5mm}\\
$\mbox{}^b$INFN, Sezione di Roma 2, Via della Ricerca Scientifica 1, 00133, Roma, Italy.\vspace{0.5mm}\\
$\mbox{}^c$INFN, Sezione di Milano, Via Celoria 16, 20133 Milano, Italy.\vspace{2mm}}}

\begin{abstract}

We present new solutions in type II supergravities describing AdS$_9$ geometries warped over an interval. In type IIB, we construct an analytic family of backgrounds supported by a non-trivial axio-dilaton profile. Despite the presence of strong-coupling singularities at both ends of the interval, these solutions exhibit both a finite Euclidean on-shell action and a finite holographic central charge. Moreover, they possess a $\mathbb Z_2$ symmetry under which the axion transforms as $C_0\rightarrow -C_0$.
We also investigate AdS$_9$ backgrounds in massive IIA supergravity supported by the Romans mass and the dilaton. Numerical integration reveals solutions with a strong-coupling singularity whose asymptotic behavior is consistent with the characteristic D8/O8 profile. In contrast to the type IIB case, our analysis indicates that the Euclidean on-shell action diverges. Finally, we identify a family of perturbative dS$_9$ solutions in massive IIA supergravity.

\end{abstract}

\maketitle

\section{Introduction}

The origin of non-supersymmetric configurations is a longstanding open question in string theory. A central idea within the swampland program is that non-supersymmetric gravitational vacua cannot be stable, as the force balance characteristic of BPS systems is generically lost in the absence of supersymmetry \cite{Arkani-Hamed:2006emk,Ooguri:2016pdq}. For a general review on the swampland program, see \cite{Palti:2019pca}.

Recently, the Cobordism Conjecture \cite{McNamara:2019rup} has significantly changed our perspective on non-BPS configurations and dynamical transitions in quantum gravity. A particularly important development along these lines is the recent proposal of domain walls connecting type IIA and type IIB theories \cite{Heckman:2025wqd,Torres:2026vxx,Anastasi:2026cus}.

An intriguing proposal in this context is the existence of R7-branes in type IIB string theory \cite{Dierigl:2022reg,Debray:2023yrs,Dierigl:2023jdp,Chakrabhavi:2025bfi}. These are non-BPS codimension-two defects associated with a reflection monodromy acting on the axio-dilaton. Recent developments suggest that they can be stable \cite{Heckman:2025wqd}.

This work is motivated by a possible connection between R7-branes and AdS$_9$ geometries, recently suggested in \cite{Cavusoglu:2026xiv} (see Appendix D therein). Constructing local spacetime realizations of non-supersymmetric configurations is a highly non-trivial problem. Progress in this direction has been achieved in the context of Dynamical Cobordism \cite{Angius:2022aeq,Angius:2023uqk,Huertas:2023syg}, as well as through bubble solutions and de Sitter-like solutions \cite{Horowitz:2007pr,Ooguri:2017njy,GarciaEtxebarria:2020xsr,Bomans:2021ara,Giri:2021eob,Dibitetto:2022rzy,Menet:2025nbf,Andriot:2026lac,Ghodsi:2026lmn}, and in non-supersymmetric string theories \cite{Dudas:2000ff,Mourad:2016xbk,Sagnotti:2021mxb,Raucci:2022bjw,Raucci:2022jgw,Mourad:2024dur,Raucci:2025bev,Basile:2026lyc}.

In this work we construct a new class of AdS$_9$ solutions in type II supergravities. Our approach is closely related to \cite{Cordova:2018eba,Cordova:2018dbb}, but in the present case analytic solutions can be obtained. Our analysis is closely related to \cite{Dibitetto:2026yft}, where we study general classes of codimension-one AdS$_D$ solutions in axio-dilaton gravity.

In type IIB, we find an analytic family of AdS$_9$ backgrounds foliated over an interval and supported by a non-trivial axio-dilaton profile. The solutions possess a $\mathbb Z_2$ symmetry under which the axion changes sign, reminiscent of the axion transformation associated with R7-branes. Although the solutions develop strong coupling singularities at the endpoints of the interval, we explicitly compute the Euclidean action and show that it is finite. This leads us to speculate that the singularities may admit a regular resolution in F-theory. Interestingly, after analytic continuation, the solutions can be interpreted as Euclidean wormholes connecting two strongly coupled regions. We also show that the holographic central charge is finite, consistent with a holographic interpretation in terms of a non-supersymmetric CFT$_8$.

We also investigate AdS$_9$ backgrounds in massive IIA supergravity. In this case, we construct numerical solutions supported by the Romans mass that interpolate between a weakly coupled region and a strongly coupled singularity. We provide numerical evidence that the asymptotic behaviour at strong coupling approaches the characteristic D8/O8 profile. Unlike the type IIB solutions, our numerical arguments suggest that the Euclidean action diverges, indicating a qualitatively different non-perturbative behaviour. Finally, we derive a family of perturbative dS$_9$ solutions in massive IIA supported by the Romans mass and the dilaton.

\section{$\text{AdS}_9$ in type IIB}

Let's consider the following geometries in type IIB supergravity in the {\itshape string frame},
\begin{equation}\label{IIBAdS9}
 ds^2_{10}=e^{2f(y)}L^2ds^2_{\text{AdS}_9}+e^{2h(y)}dy^2\,,
\end{equation}
where $ds^2_{\text{AdS}_9}$ is the metric over unit radius $\text{AdS}_9$ and $L$ is the reference length. The function $h(y)$ is pure gauge, and we will make a suitable choice for it.
We consider a non-trivial profile along the $y$-direction for the axio-dilaton $\tau=C_0+ie^{-\Phi}$, namely
\begin{equation}\label{IIBAxioDilaton}
 C_0=\chi(y)\qquad \text{and} \qquad \Phi=\phi(y)\,.
\end{equation}
This system is governed by the Einstein--Hilbert action coupled to the axio-dilaton. The equations of motion (EOM) take the form
\begin{equation}
\begin{split}
&e^{-2\Phi} \left( R_{mn} + 2 \nabla_m \nabla_n \Phi \right) - \frac{1}{2} F_{1\,m} F_{1\,n}+\frac14 g_{mn}|F_1|^2 = 0\,,\\
&4 \nabla^2 \Phi - 4 (\nabla \Phi)^2 + R = 0\,,\\
&\nabla_m \,F^m_1 =0\,,
\end{split}
\end{equation}
where the flux $F_1=dC_0$ satisfies the Bianchi identities $dF_1=0$. We can now specialize the EOM to the geometry \eqref{IIBAdS9}, obtaining the following set of second-order equations,
\begin{equation}\label{EOMIIB}
\begin{split}
&f'' =
- 9 (f')^2
+ \frac{e^{2 \phi}}{4} (\chi')^2
+ f' \left( h' + 2 \phi' \right)
- \frac{8\,e^{-2 f+ 2 h}}{L^2}\,,
\\[3pt]
&\phi'' =
- 36 \left(f'\right)^2
+ \frac{5e^{2 \phi}}{4}  \left(\chi'\right)^2
+\phi' \left(9 f' + h' \right)-\frac{36 e^{-2 f + 2 h}}{L^2},
\\[5pt]
&\chi''=
\left( -9 f' + h' \right) \chi'\,,
\end{split}
\end{equation}
where primes denote derivatives with respect to $y$.
The above equations must be supplemented by the {\itshape Hamiltonian constraint}
\begin{equation}\label{HeffIIB}
18 (f')^2 - 9 f' \phi' +  (\phi')^2 - \frac{1}{8} e^{2\phi} (\chi')^2 + \frac{18\,e^{-2f + 2h}}{L^2}=0\,.
\end{equation}

Let's now discuss the solutions of the field equations.
The system \eqref{EOMIIB} can be naturally reduced by integrating out the equation for the axion. This leads to the first-order differential condition
\begin{equation}\label{chiEq}
 \chi'=Q\,e^{-9f+h}\,,
\end{equation}
with $Q$ an integration constant.
It follows that \eqref{HeffIIB} can be rewritten as a differential constraint for $f\,,\phi$ with the following effective potential,
\begin{equation}
 V_{eff}=\frac{18}{L^2}\,e^{-2f + 2h}-\frac{Q^2}{8}\,e^{-18f + 2h+2\phi}\,.
\end{equation}
This quantity receives contributions from both the AdS$_9$ curvature and the $F_1$ flux.
In order to solve equations \eqref{EOMIIB}, we can exploit the gauge redundancy of the metric ansatz. A simple choice is the following,
\begin{equation}\label{axionEq}
 h=9f \qquad \longrightarrow \qquad \chi'=Q\,.
\end{equation}
We can now impose this choice on the EOM, together with the following reparametrization of the background,
\begin{equation} \label{EFrame}
 A= f-\frac{\phi}{4}\,, \qquad  B= h-\frac{\phi}{4}\,.
\end{equation}
We point out that this choice corresponds to casting the metric \eqref{IIBAdS9} in the {\itshape Einstein frame}.
After the gauge choice \eqref{axionEq} and rewriting the EOM in the Einstein frame, the first two equations in \eqref{EOMIIB} take the form
\begin{equation}
 \begin{split}\label{eqAphi}
&A''-2\phi'A' +\frac{8\,e^{16 A+4\phi}}{L^2}=0\,,\\[3pt]
&\phi'' -2 \left(\phi'\right)^2-Q^2\,e^{2\phi}=0\,.
\end{split}
\end{equation}
where, in order to decouple the equation for $\phi$, we used the Hamiltonian constraint \eqref{HeffIIB}.
These equations can be integrated analytically for $\phi$ and $A$.
In order to write the full solution in a simple form, we redefine the coordinate $y$ as an angular variable. The full solution can then be written as follows
\begin{equation}
 \begin{split}\label{exactAdS9}
  &e^{2A}=e^{2A_0}\,
  \cosh^{-1/4}\left(
  \frac13\,\log\left(
  c-\frac{2c}{1+\sin(Q\,\theta)}
  \right)
  \right)\,,\\[10pt]
  &e^{2B}=\frac{e^{2 A_0} L^2 Q^2 \cos^{-2}\left(Q\, \theta\right)}
{144 \cosh^{9/4}\!\left(
\frac{1}{3}\log\!\left(
c - \frac{2 c}{1 + \sin\bigl(Q \,\theta \bigr)}
\right)
\right)}\,,\\[10pt]
  &e^\phi=
  \frac{Q\,L\,e^{-8A_0}}
  {12\,\cos(Q\,\theta)}\,,\\[10pt]
  &\chi=
  \chi_0+\frac{12\,e^{8A_0}\sin(Q\,\theta)}{L\,Q}\,,
 \end{split}
\end{equation}
where $\theta \equiv y-y_0$ and $(A_0, Q, y_0, c,\chi_0)$ are integration constants.  This parametrization is introduced in the general classification presented in Sec. 3.2 in \cite{Dibitetto:2026yft}. We can now study in detail the main properties of this solution.

\subsection{The study of the solution}

The solution is completely real for $c<0$ and regular in the intervals $y\in  (y_0-\frac{\pi(1-2n)}{2Q},\, y_0+\frac{\pi(1+2n)}{2Q})$. Each interval is centered around $y_0$ and adjacent intervals are separated by singular loci where the dilaton diverges.

Let us now discuss the flux associated with the axion field. Integrating $F_1=d\chi$ over a regular interval yields the following result
\begin{equation}
N= \frac{1}{2\pi}\int F_1
=  \frac{12\,e^{8A_0}}{\pi LQ}\,.
\end{equation}
The appearance of a non-trivial integrated $F_1$ flux suggests a possible relation to quantum objects carrying SL$(2,\mathbb{Z})$ charges. However, the precise D-brane interpretation associated with this quantity is not immediately clear from our solution.

In what follows, we discuss several arguments that may help clarify the quantum gravity origin of these solutions. First of all, we point out that \eqref{exactAdS9} possesses a $\mathbb{Z}_2$ symmetry under reflection of the $y$ coordinate. Indeed, under the spacetime reflection $\theta \rightarrow -\theta$, the metric and the dilaton in \eqref{exactAdS9} remain invariant, while, up to a constant shift, the axion transforms as
\begin{equation}
 C_0 \rightarrow -C_0\,.
\end{equation}
We note that this transformation of the axion is reminiscent of the reflection monodromy discussed in \cite{Dierigl:2022reg,Debray:2023yrs,Dierigl:2023jdp,Chakrabhavi:2025bfi,Cavusoglu:2026xiv} in the context of R7-branes.

\begin{figure}[H]
\centering
\includegraphics[width=0.7\columnwidth]{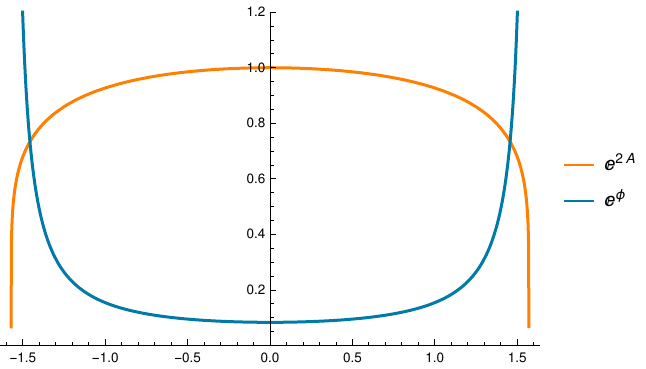}
\caption{An AdS$_9$ solution with $L=1,\, A_0=0,\, Q=1,\, y_0=0, \, c=-1$. The solution is plotted in the interval $y \in (-\frac{\pi}{2}, \frac{\pi}{2})$.}
\label{fig:AdS9Aphi}
\end{figure}

\begin{figure}[H]
\centering
\includegraphics[width=0.7\columnwidth]{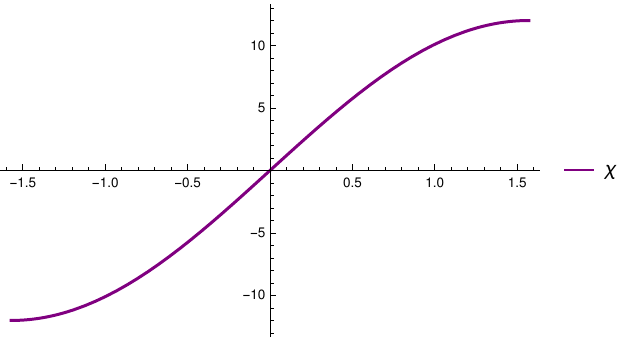}
\caption{Profile of the axion field $C_0=\chi$ for the same solution depicted in Figure \ref{fig:AdS9Aphi}.}
\label{fig:AdS9C0}
\end{figure}

Further insight can be gained by studying the behavior of the solution near the endpoints of the interval. Let us present the solution for a representative choice of parameters. An example is shown in Figures \ref{fig:AdS9Aphi} and \ref{fig:AdS9C0}.

From the above figure, we can explicitly see that the dilaton diverges at both endpoints. In these limits, the solution becomes strongly coupled and the ten-dimensional supergravity description is no longer reliable. This naturally raises the question of whether the endpoint singularities are intrinsic to the solution or instead admit a higher-dimensional resolution.
An analogous situation is discussed in Sec. 3.1 in \cite{Dibitetto:2026yft} for AdS$_7$ solutions of eight-dimensional SU$(2)$ gauged supergravity. Although the eight-dimensional description exhibits a diverging dilaton, the solution admits a regular uplift to the non-supersymmetric AdS$_7\times S^4$ vacuum of M-theory \cite{Salam:1984ft}. In the uplifted geometry, the singular endpoints of the interval are mapped to the two poles of the internal $S^4$, yielding a completely smooth eleven-dimensional background. From the lower-dimensional perspective, these singularities can be interpreted as KK monopoles filling AdS$_7$, whose apparent singular behavior is entirely resolved in eleven dimensions.

Motivated by this example, a natural question is whether our AdS$_9$ solutions admit a similar resolution. A first indication comes from the finiteness of the on-shell action. Further evidence in this direction is provided by the fixed point analysis of Secs.~2.2 and 3.2 in \cite{Dibitetto:2026yft}.

First of all, we note that our solutions admit an interesting interpretation as Euclidean wormholes connecting two strongly coupled regions. Indeed, after a suitable analytic continuation of the time direction, the metric \eqref{IIBAdS9} can be rewritten as a regular foliation of the hyperbolic plane $\mathbb{H}^9$. In this way, one obtains a co-dimension one Euclidean solution exhibiting a $\mathbb Z_2$ symmetry along the foliation coordinate.

This interpretation of our solutions as gravitational instantons can be further supported by the explicit derivation of the on-shell action. For this computation, we refer to \cite{Dibitetto:2020csn}, where the derivation is carried out for a general class of warped backgrounds of the type \eqref{IIBAdS9}. In general, the action consists of a bulk contribution, namely the standard type IIB action associated with the EOM \eqref{EOMIIB}, together with the Gibbons--Hawking--York (GHY) term accounting for boundary contributions. One can verify that the bulk action vanishes on-shell, so that the entire contribution arises from the GHY term. In our case, the boundary is identified with the two endpoints of the interval $I=(y_0-\frac{\pi}{2Q}, \,y_0+\frac{\pi}{2Q})$.

In Section~3.1 of \cite{Dibitetto:2020csn}, the computation of the GHY term is presented in detail. Applying that analysis to the present case yields the on-shell action
\begin{equation}\label{GHY}
S_{E, \text{GHY}}=-\frac{9 L^9\mathrm{Vol}(\mathbb{H}^9)}{8\pi G_N}\,e^{9A-B}A'\bigr|_{\partial I}
=\frac{9L^8 e^{8A_0}\mathrm{Vol}(\mathbb{H}^9)}{4G_N},
\end{equation}
where $G_N$ is the 10D Newton constant and $\mathrm{Vol}(\mathbb{H}^9)$ is the regularized volume of the hyperbolic space $\mathbb{H}^9$.

Since genuinely pathological singularities are generally expected to lead to divergent physical observables, a finite on-shell action motivates a higher-dimensional interpretation of the endpoint singularities. One may speculate that these solutions admit a resolution within F-theory, although no local uplift is currently known. Developing these ideas explicitly constitutes a very interesting open problem, which is likely related to understanding the non-perturbative origin of our AdS$_9$ solutions.

Finally, a complementary computation is the holographic central charge for the AdS$_9$ solutions \eqref{exactAdS9}. Using the standard formulas of \cite{Klebanov:2007ws}, one obtains
\begin{equation}
c_{\text{hol}}=\frac{3}{2^7\,G_N}\,\int dy\,e^{7A+B}\,.
\end{equation}
Unfortunately, we could not find a closed-form expression for this integral. Nevertheless, using the explicit solution \eqref{exactAdS9}, one can still show that the holographic central charge is finite. For $Q>0$ and $c<0$, the integrand entering the holographic central charge is continuous on each regular interval. Moreover, one can explicitly verify that it vanishes at both endpoints. Therefore, the integrand extends continuously to the corresponding closed domain and, by the Weierstrass theorem, it is necessarily bounded. This implies that the holographic central charge is finite.

From this analysis it follows that, although we could not determine the universal large-$N$ scaling of $c_{\text{hol}}$, the finiteness of the holographic free energy is consistent with a possible holographic interpretation of our backgrounds in terms of a non-supersymmetric CFT$_8$.

\subsection{Perturbative solutions}\label{sec:pertIIB}

In this section we study AdS$_9$ solutions in type IIB by expanding the metric and the axio-dilaton about a regular point, and by solving the EOM order by order in $y$. Unlike the analytic solution discussed previously, we will not choose a particular gauge for the function $h$.

Let's start from the string frame metric \eqref{IIBAdS9} and from the prescription \eqref{IIBAxioDilaton} for the axio-dilaton. We consider the following expansion of the fields about the $y=0$ region,
\begin{equation}
 \begin{split}\label{PertExpanIIB}
&f=f_0+f_1\,y+f_2\,y^2+f_3\,y^3+f_4\,y^4+O(y^5)\,,\\
&h=h_0+h_1\,y+h_2\,y^2+h_3\,y^3+h_4\,y^4+O(y^5)\,,\\
&\phi=\phi_0+\phi_1\,y+\phi_2\,y^2+\phi_3\,y^3+\phi_4\,y^4+O(y^5)\,.
 \end{split}
\end{equation}
As for the axion $\chi$, we impose the first-order condition \eqref{chiEq} and reduce \eqref{EOMIIB} to a second-order system for $f$ and $\phi$. We thus obtain a system of two second-order equations, together with the first-order Hamiltonian constraint \eqref{HeffIIB}. We can evaluate these equations on the perturbative expansion \eqref{PertExpanIIB} and solve them order by order in $y$. We obtain the following solution,
\begin{equation}
\begin{split}\label{pertSolIIB}
&\phi_2=
\frac{
72 \sqrt{2}\,3^{1/4} e^{2 h_0-\frac{\phi_0}{4}}
}{
L^{9/4}Q^{1/4}
}\,,
\qquad
h_2=
\frac{
118 \sqrt{2}\,3^{1/4} e^{2 h_0-\frac{\phi_0}{4}}
}{
5 L^{9/4}Q^{1/4}
}\,,\\[0.7em]
&f_2=
\frac{
14 \sqrt{2}\,3^{1/4} e^{2 h_0-\frac{\phi_0}{4}}
}{
L^{9/4}Q^{1/4}
}\,,
\qquad
f_4=
-\frac{
32 e^{4 h_0-\frac{\phi_0}{2}}
}{
5 \sqrt{3}\,L^{9/2}\sqrt{Q}
}\,,\\[0.7em]
&h_4=
-\frac{
391136 e^{4 h_0-\frac{\phi_0}{2}}
}{
75 \sqrt{3}\,L^{9/2}\sqrt{Q}
}\,,
\qquad
\phi_4=
\frac{
2688 \sqrt{3}\,e^{4 h_0-\frac{\phi_0}{2}}
}{
5 L^{9/2}\sqrt{Q}
}\,,\\[0.7em]
&f_0=
\frac{\phi_0}{8}
+\frac{1}{8}\log\left(
\frac{L\,Q}{12}
\right)\,,
\end{split}
\end{equation}
where the odd terms vanish, $h_1=f_1=\phi_1=h_3=f_3=\phi_3=0$. We were able to extend this solution up to $O(y^6)$ order, finding that $f_5=h_5=\phi_5=0$ and non-trivial values for $f_6,\, h_6,\, \phi_6$. For the sake of simplicity, we omit the explicit form of the solution at this order. Interestingly, by extending the perturbative integration to order $O(y^6)$, one finds an additional free parameter that first appears at order $O(y^4)$. In other words, the perturbative expansion \eqref{pertSolIIB} belongs to a broader parametric family of type IIB solutions.

The question now is whether the exact solutions \eqref{exactAdS9} and the perturbative ones presented in this section belong to the same class. Providing a precise answer to this question is not straightforward. Nevertheless, we can discuss several arguments suggesting that this could indeed be the case. To this aim, let us expand the analytic solution \eqref{exactAdS9} around $y=0$ with $y_0=0$. One immediately observes that the expansion also contains odd powers in $y$. We point out that this fact does not provide very deep information, since the analytic solution was obtained after choosing the very particular parametrization \eqref{axionEq}. However, if we impose $c=-1$ in \eqref{exactAdS9}, all odd powers disappear and the expansion of $f$, $h$ and $\phi$ takes the same form as \eqref{PertExpanIIB}. This fact is promising. The solution \eqref{exactAdS9} around $y=0$ with $c=-1$ takes the local form
\begin{equation}
\begin{split}\label{AdS9expandedExact}
&f =
\frac{1}{4}\log\!\left(\frac{e^{-4A_0}LQ}{12}\right)
+\frac{7Q^2}{72} y^2
+\frac{53\,Q^4}{3888} y^4
+O(y^5)\,,
\\[4pt]
&h=
\frac{1}{4}\log\!\left(\frac{e^{-4A_0}(LQ)^5}{12^5}\right)
+\frac{3Q^2}{8}\,y^2
+\frac{17Q^4}{432}\,y^4
+O(y^5)\,,
\\[4pt]
&\phi =
\log\!\left(\frac{e^{-8A_0}LQ}{12}\right)
+\frac{Q^2}{2} y^2
+\frac{Q^4}{12} y^4
+O(y^5)\,.
\end{split}
\end{equation}
We can try to compare this local solution with the perturbative one in \eqref{PertExpanIIB}. The comparison is non-trivial, since the two solutions are written in different parametrizations of the transverse coordinate, namely $\tilde y=\tilde y(y)$, with $e^{h(y)}dy=e^{\tilde h(\tilde y)}d\tilde y$. Moreover, we do not know the general relation between $h$ and the remaining variables underlying the perturbative solution \eqref{PertExpanIIB}.

Nevertheless, the solutions for $f$ and $\phi$ remarkably match at both $O(y^0)$ and $O(y^2)$. Indeed, at zeroth order, the combination $f-\frac{\phi}{8}$ in \eqref{AdS9expandedExact} precisely reproduces the relation between $f_0$ and $\phi_0$ appearing in \eqref{PertExpanIIB}. The agreement persists at quadratic order. In particular, one can show that the second-order coefficients in \eqref{PertExpanIIB} coincide with the corresponding ones in \eqref{AdS9expandedExact}, namely $f_2=f^{ex}_2$ and $\phi_2=\phi^{ex}_2$, where $f^{ex}_2=\frac{7Q^2}{72}$ and $\phi^{ex}_2=\frac{Q^2}{2}$ are extracted from \eqref{AdS9expandedExact}. The situation changes at quartic order, where the corresponding coefficients no longer agree.

This behavior can be understood from reparametrization invariance. If the two backgrounds belong to the same local branch of solutions, they should be related by a regular local redefinition of the radial coordinate,
\begin{equation}
\begin{split}\label{repar}
 &\tilde y=a_1\,y+a_3\,y^3+a_5\,y^5+\dots\,,\\
& X(\tilde y)=X_0+a_1^2X_2\,y^2+\left(2a_1a_3X_2+a_1^4X_4\right)y^4+\dots\,,
\end{split}
\end{equation}
where $X(\tilde y)=X_0+X_2\tilde y^2+X_4\tilde y^4+\dots$ is the expansion of a generic function defined only by even powers of $\tilde y$. One can then explicitly see that the zeroth-order coefficients are invariant, while the quadratic coefficients are only rescaled by an overall factor $a_1^2$. In the present case, the matching of the quadratic coefficients implies $a_1=1$. Therefore, the first possible mismatch between the two parametrizations can only arise through the cubic coefficient $a_3$.

This coefficient can be extracted from the expansion of the function $h$. Under the redefinition \eqref{repar}, one has $h(y)=\tilde h(\tilde y)+\log\left(\frac{d\tilde y}{dy}\right)$, where the tilded quantities correspond to the solution \eqref{AdS9expandedExact}. For $a_1=1$, this implies $h_0=\tilde h_0$ and $h_2=\tilde h_2+3a_3$. Using the explicit solutions, one then finds $a_3=-\frac{19Q^2}{270}$. Therefore, the quadratic coefficient of $h$ is already sensitive to the cubic correction in the coordinate transformation.

Indeed, once $a_3\neq0$, the quartic coefficients in \eqref{repar} receive additional contributions proportional to the lower-order term $X_2$. Therefore, quartic coefficients are no longer invariant under local redefinitions of the transverse coordinate. This explains why the matching observed at quadratic order is not expected to persist automatically at $O(y^4)$ order. However, as previously mentioned, the perturbative solution develops an additional free parameter precisely at quartic order when $O(y^6)$ contributions are included in the solution. This further suggests that the mismatch may be related to residual parametrization freedom rather than to a genuine inequivalence between the two local solutions.

\section{$\text{AdS}_9$ in massive IIA}

Let's now consider AdS$_9$ solutions of the type \eqref{IIBAdS9} in the context of massive IIA supergravity. We may recall here the prescription for the {\itshape string frame} metric and the dilaton,
\begin{equation}\label{IIAAdS9}
\begin{split}
& ds^2_{10}=e^{2f(y)}L^2ds^2_{\text{AdS}_9}+e^{2h(y)}dy^2\,,\\
&\Phi=\Phi(y)\,.
\end{split}
\end{equation}
Instead of the axion contribution, we search for solutions with a geometry supported by the Romans mass $F_0=m$. The equations of motion take the following form
\begin{equation}
\begin{split}
&e^{-2\Phi} \left( R_{mn} + 2 \nabla_m \nabla_n \Phi \right) +\frac14 g_{mn}|F_0|^2 = 0\,,\\
&4 \nabla^2 \Phi - 4 (\nabla \Phi)^2 + R = 0\,.
\end{split}
\end{equation}
Following the same procedure as in the previous section, we can evaluate these equations on the AdS$_9$ backgrounds \eqref{IIBAdS9}. We obtain
\begin{equation}\label{EOMIIA}
\begin{split}
&f'' =
- 9 (f')^2
+ f' \left( h' + 2 \phi' \right)
- \frac{8e^{-2 f+ 2 h}}{L^{2}}+\frac{m^2e^{2 h+ 2 \phi}}{4} \,,
\\[3pt]
&\phi''=
- 36 (f')^2 + \phi'(9f'+h')
- \frac{36e^{-2 f+ 2 h}}{L^{2}}+ m^2 e^{2 h+ 2 \phi}\,.
\end{split}
\end{equation}
The Hamiltonian constraint takes the following form,
\begin{equation}\label{HeffIIA}
18 (f')^2 - 9 f' \phi' +  (\phi')^2+ \frac{18e^{-2f + 2h}}{L^{2}}+\frac{m^2 e^{2 h+ 2 \phi}}{8} =0\,.
\end{equation}

Unfortunately, in this case we were not able to find an analytic solution as in the type IIB setup. Nevertheless, in what follows we numerically integrate the equations of motion.
In order to maintain the analogy with the type IIB solutions, we cast the system in the Einstein frame \eqref{EFrame}, namely $f=A+\frac{\phi}{4}$ and $h=B+\frac{\phi}{4}$.
First of all, we need to fix a parametrization for the $y$-coordinate, namely we have to make a choice for $B$. A particularly simple choice is the conformal gauge $B=A$.
\begin{figure}[H]
\centering
\includegraphics[width=0.7\columnwidth]{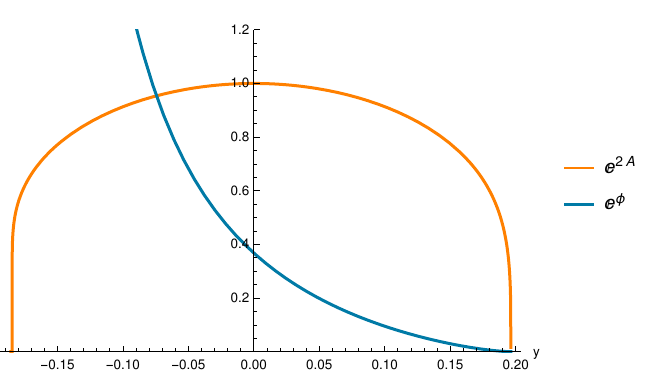}
\caption{An AdS$_9$ solution in massive IIA. The region where $e^{\Phi}\rightarrow +\infty$ reproduces a D8/O8 singularity.}
\label{fig:AdS9AphiIIA}
\end{figure}
Before performing the numerical integration, we need to choose the initial conditions for the fields $A$, $\phi$, and for their derivatives. An example of a numerical solution is shown in Figure \ref{fig:AdS9AphiIIA}. The shooting procedure starts at $y=0$ with $A(0)=0$, $A'(0)=0$, $\phi(0)=-1$, and $\phi'(0)\simeq -12.006$. The last value is obtained by solving the Hamiltonian constraint \eqref{HeffIIA} with $L=1$ and $m=1$. In fact, since we are solving a system of two second-order equations, \eqref{EOMIIA}, supplemented by the first-order differential constraint \eqref{HeffIIA}, the initial conditions must be chosen consistently with the constraint itself.
As a result, we obtain the solution depicted in Figure \ref{fig:AdS9AphiIIA}. The functions $A$ and $\phi$ interpolate between two singular regions located at $y_-\simeq-0.184$ and $y_+\simeq 0.196$. At the singularity $y=y_+$ the solution remains weakly coupled. On the other hand, as $y\rightarrow y_-$ the background becomes strongly coupled.

Comparing these solutions with those in Figure \ref{fig:AdS9Aphi}, we observe a qualitative difference. In type IIB we proposed that the strongly coupled regions could be smoothly resolved in a twelve-dimensional geometry. In the present case, however, the situation is rather different. Since vacuum configurations in massive IIA are associated with strongly coupled D8/O8 sources, it is natural to expect that the regions where $e^{\Phi}\rightarrow +\infty$ correspond to a stack of D8/O8 branes extended along the AdS$_9$ slices. We point out that O8 singularities may be associated with tachyonic instabilities, as described in \cite{Bena:2020qpa}. Whether a similar mechanism is present in the solutions discussed here remains an open question.

\begin{figure}[H]
\centering
\includegraphics[width=0.49\columnwidth]{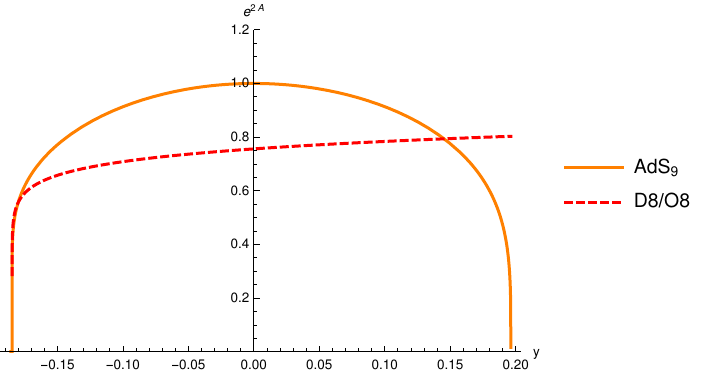}
\hfill
\includegraphics[width=0.49\columnwidth]{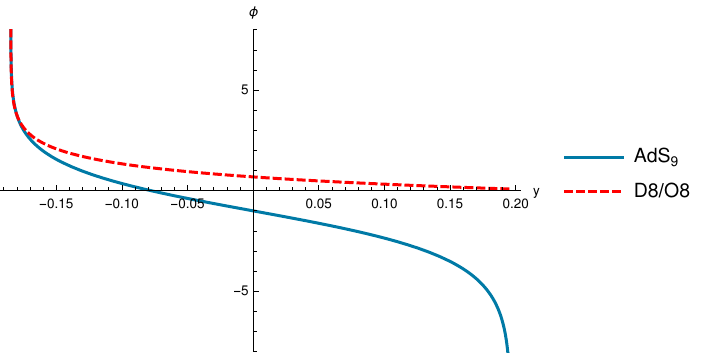}
\caption{Comparison between the numerical AdS$_9$ solution and the D8/O8 solution near the strongly coupled singularity at $y=y_-$. The left panel shows $e^{2A}$, while the right panel shows $\phi$.}
\label{fig:AdS9D8}
\end{figure}

In order to check this behavior, let us recall the local form of a D8/O8 singularity in the Einstein frame. In the conformal gauge $B=A$ adopted for the numerical solutions, the metric and the dilaton behave as
\begin{equation}
\begin{split}\label{D8}
&e^{2A}\sim(y-y_{\text{D}8})^{1/12}\,,\qquad \quad e^{\Phi}\sim(y-y_{\text{D}8})^{-5/6}\,,
\end{split}
\end{equation}
where $y_{\text{D}8}$ denotes the position of the D8/O8 source. We can therefore compare these asymptotics with the numerical solutions near the strongly coupled region $y\rightarrow y_-$. In Figure \ref{fig:AdS9D8} we observe that the numerical profiles of both $e^{2A}$ and $\phi$ are well approximated by the D8/O8 behavior close to the singularity.
The D8/O8 profiles shown in Figure \ref{fig:AdS9D8} have been obtained by comparing the AdS$_9$ solution with the behaviors \eqref{D8}. The position of the D8/O8 source has been identified with the strongly coupled endpoint of the numerical solution, namely $y_{\text{D}8}=y_-\simeq -0.184$. The overall normalization of the asymptotic curves has then been fixed by requiring agreement with the numerical profiles at a nearby reference point close to the singularity. For a more detailed discussion of the asymptotic behavior of these solutions from the perspective of the fixed point analysis, see Sec.~3.4 of \cite{Dibitetto:2026yft}.

Finally, we discuss the Euclidean on-shell action associated with the massive IIA numerical backgrounds. In this case, the numerical solution suggests that the on-shell action is not finite due to the presence of a D8/O8 singularity at $y=y_-$. To investigate this issue, let us focus again on the GHY contribution.
Adapting the general expression \eqref{GHY} to the conformal gauge $B=A$, one obtains
\begin{equation}\label{GHYIIA}
S_{E,\mathrm{GHY}}
=
-\frac{9L^9\mathrm{Vol}(\mathbb H^9)}{8\pi G_N}
\,\left[(e^{8A}A')|_{y=y_+}-(e^{8A}A')|_{y=y_-}\right]\,.
\end{equation}
In the numerical analysis we factor out $\mathrm{Vol}(\mathbb H^9)$. Moreover, we set $L=1$ and $8\pi G_N=1$.

\begin{figure}[H]
\centering
\includegraphics[width=0.7\columnwidth]{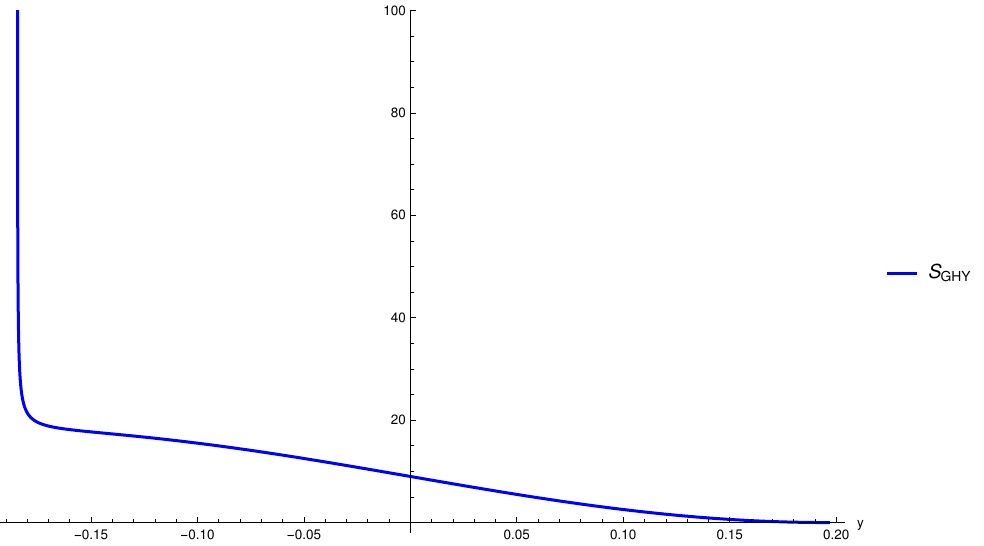}
\caption{The GHY action for the AdS$_9$ solution in massive IIA supergravity. The action is evaluated on $[y,y_+]$ and grows rapidly near the D8/O8 singularity at $y=y_-$.}
\label{fig:AdS9IIAGHY}
\end{figure}
To obtain Figure~\ref{fig:AdS9IIAGHY}, we fix the endpoint $y_+$ and evaluate the GHY contribution on the interval $[y,y_+]$.
The resulting profile shows that the GHY contribution increases rapidly as the D8/O8 singularity is approached and eventually diverges. This behavior indicates that the Euclidean action is not finite for the massive IIA solutions considered here.

\subsection{Perturbative dS$_9$ solutions}

As mentioned above, we were not able to find analytic AdS$_9$ solutions in massive IIA supergravity. Nevertheless, this theory admits a class of perturbative dS$_9$ solutions. These are obtained by integrating the EOM order by order, following the same approach adopted for the type IIB solutions in \eqref{pertSolIIB}. Although little can be said about the global properties of these solutions, they nevertheless suggest, as in the AdS$_9$ case, an interpretation as gravitational instantons in massive IIA supergravity. Let us consider the following background
\begin{equation}\label{dS9}
 ds^2_{10}=e^{2f(y)}L^2ds^2_{\text{dS}_9}+e^{2h(y)}dy^2\,,
\end{equation}
where $ds^2_{\text{dS}_9}$ is the metric on unit radius de Sitter space. In addition to the metric, we include a running dilaton $\Phi=\phi(y)$ and the Romans mass.

As in the type IIB case, we expand all fields around $y=0$ as
\begin{equation}
 \begin{split}\label{PertExpanIIA}
&f=f_0+f_1\,y+f_2\,y^2+f_3\,y^3+f_4\,y^4+O(y^5)\,,\\
&h=h_0+h_1\,y+h_2\,y^2+h_3\,y^3+h_4\,y^4+O(y^5)\,,\\
&\phi=\phi_0+\phi_1\,y+\phi_2\,y^2+\phi_3\,y^3+\phi_4\,y^4+O(y^5)\,.
 \end{split}
\end{equation}
We then evaluate the equations of motion and the Hamiltonian constraint on the perturbative expansion \eqref{PertExpanIIA}, and solve them order by order in $y$. This yields
\begin{equation}
\begin{split}
\phi_2&=
\frac{5}{8}\,m^2 e^{2(h_0+\phi_0)}\,,
\qquad
h_2=
-\frac{59}{360}\,m^2 e^{2(h_0+\phi_0)}\,,
\\[0.5em]
f_2&=
\frac{11}{72}\,m^2 e^{2(h_0+\phi_0)}\,,
\qquad
f_4=
\frac{61}{38880}\,m^4 e^{4(h_0+\phi_0)}\,,
\\[0.5em]
h_4&=
-\frac{64271}{583200}\,m^4 e^{4(h_0+\phi_0)}\,,
\qquad
\phi_4=
\frac{31}{864}\,m^4 e^{4(h_0+\phi_0)}\,,
\\[0.5em]
f_0&=
-\phi_0+
\log\!\left(
\frac{12}{Lm}
\right)\,.
\end{split}
\label{pertSolIIA}
\end{equation}
where the odd coefficients vanish, namely $h_1=f_1=\phi_1=h_3=f_3=\phi_3=0$. Finally, we observe that by extending the perturbative solution to order $O(y^6)$, one finds an additional free parameter which is not fixed by the EOM. This suggests the existence of a broader family of dS$_9$ solutions whose global properties deserve further investigation.
\newline
\noindent {\bf Acknowledgements:}
 N.P. thanks the members of the hep-th group at the University of Turin for their kind hospitality while parts of this work were carried out.

\balance

\bibliographystyle{apsrev4-2}
\bibliography{references}

\end{document}